\documentclass[useAMS,usenatbib]{mn2e}
\usepackage{graphicx}

\usepackage{txfonts}

\title[Radio emission from hot magnetic stars]{First detections of 610 MHz radio emission from hot magnetic stars}
\author[Chandra et al.]{P.~Chandra$^{1}$\thanks{E-mail: poonam@ncra.tifr.res.in}, 
G.~A.~Wade$^{2}$, J.~O.~Sundqvist$^{3}$,  D.~Oberoi$^1$, 
J.~H.~Grunhut$^4$,
A.~ud-Doula $^{5}$,
\newauthor
 V.~Petit$^{6}$, D.~H.~Cohen$^{7}$,
M.~E.~Oksala$^{8}$,
and A.~David-Uraz$^{2,9}$\\
$^{1}$National Centre for Radio Astrophysics, Tata Institute of Fundamental Research, P.O. Box 3, 
Pune 411007, India\\
$^{2}$Department of Physics, Royal Military College of Canada, PO Box 17000, Station Forces, Kingston, Ontario K7K 7B4, Canada\\
$^{3}$Department of Physics \& Astronomy, University of Delaware,  Newark, DE 19716, USA\\
$^{4}$European Southern Observatory, Karl-Schwarzschild-Str. 2, D-85748 Garching, Germany\\
$^5$Penn State Worthington Scranton, 120 Ridge View Drive, Dunmore, PA 18512, USA \\
$^6$Dept. of Physics \& Space Sciences, Florida Institute of Technology,
Olin Physical Science, 346, Melbourne, FL 32901, USA\\
$^7$Department of Physics and Astronomy, Swarthmore College, Swarthmore, PA 19081, USA\\
$^8$LESIA, Observatoire de Paris, CNRS UMR 8109, UPMC, Universit\'{e} Paris Diderot, 5 place Jules Janssen, 92190 Meudon, France\\
$^9$Department of Physics, Engineering Physics and Astronomy, Queen's
University, 99 University Avenue, Kingston, ON K7L 3N6, Canada
} 
\begin{document}

\date{Submitted 8th May 2015}

\pagerange{\pageref{firstpage}--\pageref{lastpage}} \pubyear{2000}

\maketitle

\label{firstpage}

\begin{abstract}
We have carried out a study of  radio emission from a 
small sample of magnetic O- and B-
type stars using the Giant Metrewave Radio Telescope, 
with the goal of investigating their
magnetospheres at low frequencies. 
These are the lowest frequency radio measurements ever obtained 
of hot magnetic stars. 
The observations were taken at random rotational
phases in the 1390 and the 610 MHz bands. Out of the 8 stars, we detect  
five B-type stars in both the 1390
and the 610 MHz bands. The O-type stars were
observed only in the 1390 MHz band, and no detections were obtained.
 We explain this 
result as a consequence of free-free absorption by the free-flowing stellar
wind exterior to the closed magnetosphere.
We also study the variability of individual stars. One star - HD 133880 - exhibits remarkably strong and rapid variability 
of its low frequency flux density. We discuss the possibility of this
emission being coherent emission as reported for CU Vir by \citet{trigilio00}.

\end{abstract}


\section{INTRODUCTION}

The last decade has witnessed the identification and elaboration of an important 
population of hot O- and B-type stars hosting strong, organized magnetic fields 
\citep[e.g.][]{donati02, oksala10, grunhut12, wade12}. Observational and theoretical 
studies of the interaction of their intense radiation-driven winds
with their magnetic fields \citep[e.g.][]{uddoula02,2005ApJ...630L..81T,sund12} have
revealed that the stellar wind properties of magnetic hot stars are modified
in important ways as compared to the winds of non-magnetic stars,
introducing significant changes across the entire electromagnetic spectrum.

The presence of an organized magnetic field at the surface of a hot star leads to channelling and confinement of its outflowing
stellar wind, creating a magnetosphere, which can radiate in various wavebands 
\citep{andre88,linsky92}.

In regions close to the star,  the magnetic pressure dominates and causes the wind to follow dipolar field lines. At greater distances from the stellar surface, the wind kinetic energy density
 exceeds the magnetic pressure due to the stronger radial decline of magnetic field energy density than of wind kinetic energy density. 
The radius at which  
the wind kinetic energy density
becomes equal to the magnetic pressure is defined
as  the Alfv\'en radius \citep[e.g.][]{uddoula02}.
The region below the Alfv\'en radius, i.e.  interior to the
largest closed magnetic loop,
is defined as the inner magnetosphere \citep{trigilio04}, which is the site of generation of
X-ray and H$\alpha$ emission
 \citep[e.g. ][]{bm97,gagne05,howarth07,sund12}. 
Beyond the Alfv\'en radius, the wind opens the magnetic field lines, generating a 
current sheet in the magnetic equatorial plane. This region is defined as the
middle magnetosphere.
At greater distances from the star, the outer magnetospheric region is characterized by a radial magnetic topology following wind streamlines.

Radio emission from non-magnetic hot stars is expected to be thermal
free-free emission from the ionized wind in the circumstellar environment \citep{pf75,wb75}.
However, in the presence of magnetic fields
electrons can be accelerated to relativistic energies, either by 
reconnection near the current sheet in the middle magnetosphere \citep{um92} or generated in strong, large-scale shocks in the inner magnetosphere \citep{bm97,uto06,uot08}. These energetic electrons can give rise to 
gyrosynchrotron radio emission \citep{drake87,linsky92,trigilio04}.

Various surveys of cooler magnetic A- and B-type stars have been carried out at 5--8 GHz radio 
frequencies, resulting  
 in $\approx 25\%$ detections \citep[e.g. ][]{abott85,drake87,linsky92}. \citet{linsky92} interpreted the radio emission in these stars
as gyrosynchrotron in nature, arising
from regions located at $10-20~R_\star$, with higher frequency emission 
originating from plasma located closer to the star, and the lower frequency
emission from material further away. 

\citet{lo93} monitored the 5 GHz radio emission of
two magnetic B-type stars, HD 37017 and $\sigma$ Ori E, 
and observed variability in accordance with their respective rotational periods.
The coincidence of radio maxima with the
extrema of the longitudinal magnetic field in these stars 
led those authors to
suggest that radio emitting regions are located above the magnetic poles. 
Radio variability was also measured from the magnetic B8p star
HD 133880 (HR Lup). The flux density and circular polarization 
were observed to vary according to the stellar rotation period \citep{lim96,bailey12}, and was also interpreted as gyrosynchrotron emission \citep{bailey12}.

\begin{table*}
 \centering
  \caption{Details (spectral type, effective temperature, luminosity, radius, mass, rotational period, polar strength of the magnetic dipole at the stellar surface, Kepler co-rotation radius, Alfv\'en radius,
  mass loss rate and wind terminal velocity) of magnetic O and B stars observed with the GMRT. 
$R_{\rm ff}$ is the radius of the free-free absorption photosphere at 1390~MHz, discussed in Sect.~\ref{freefree}. \label{tab:stars}}
  \begin{tabular}{llcrrrrrrrrrrrrrrrr}
  \hline
   Star & Spectral &  $T_{\rm eff}$  & $\log L_\star$ &  $R_\star$ & $M_\star$ & Period & $B_{\rm p}$ & $R_{\rm K}$ & $R_{\rm A}$ & $\log(\dot M)$ & 
$v_{\infty}$ & $R_{\rm ff}$ &  $R_{\rm ff}/R_{\rm A}$\\
  & Type &  (kK) &($L_\odot$) & ($R_\odot$) & ($M_\odot$) & (d) & (kG) & ($R_*$) & ($R_*$) &($M_\odot$ yr$^{-1}$) & (km s$^{-1}$) & ($R_*$) & \\
 \hline
HD215441 &  B8-9p &  $15\pm2$ & 1.8 & 2.6 & 3.5 & 9.49  & 34 &          -   &   -&-&- & - & - \\
HD37479 &  B2 Vp & $23\pm2^1$ & 3.6$^1$ & 3.9$^1$ & 5.0$^1$
 & 1.19$^2$ & 9.6$^3$ & 2.1$^3$ & 31$^3$ &  $-9.774^6$ &1794$^6$ &   6.5&  0.2\\
HD37017 &  B2 Vp & $21\pm2^4$ &3.4$^4$ & 3.9$^4$ & 7.2$^4$ & 
0.90$^4$ & $>6.0^5$ & 1.9$^6$ & $>18^6$ & $-9.072^6$ & 1102$^6$&   27.6& $<1.5$\\
HD36485 &  B3 Vp & $20\pm2^7$ & 3.5$^7$ & 4.5$^7$ & 7.1$^7$
 & 1.48$^7$ & 10$^7$ & 2.4$^6$ & 24$^6$&  $-8.969^6$ & 1012$^6$&     30.4& 1.3\\
HD133880 &  B8IVp & $13\pm1^8$ & 2.0$^8$ & 2.0$^8$ & 3.2$^8$ & 
0.88$^8$& 19$^8$ & 2.2$^8$ & 60$^8$&   $-11.0^8$ & 750$^8$ &          4.6   &   0.1 \\
HD37022 &  O7 Vp & $39\pm1^9$ & 5.3$^9$ & 9.9$^9$ & 45$^9$
 & 15.42$^{10}$ & 1.1$^{11}$ & 9.4$^6$ & 2.4$^6$& $-6.399^6$ & 3225$^6$&    236.3& 98.5\\
HD57682 & O9 V & $34\pm1^{12}$ & 4.8$^{12}$ & 7.0$^{12}$ & 17$^{12}$ & 63.57$^{13}$ & 1.7$^{13}$ & 24$^{6}$ & 3.7$^{6}$& $-7.079^6$& 2395$^6$&              153.7& 41.5\\
NGC1624-2  & O6.5f?cp& $35\pm2^{14}$ & 5.1$^{14}$ & 9.7$^{14}$ & 34$^{14}$
 & 158.0$^{14}$ & $>20^{14}$ & 41$^{14}$ & $>11^{14}$&  $-6.786^6$&2890$^6$&      151.2& $<13.7$\\
\hline
\end{tabular}
 $^1$Hunger et al. 1989, $^2$Townsend et al. 2010,
$^3$Oksala et al. 2012, $^4$Bolton et al. 1998, $^5$Bohlender
et al. 1987, $^6$Petit et al. 2013, $^7$Leone et al. 2010, 
$^8$Bailey et al. 2012, $^{9}$Simon-Diaz et al. 2006,
 $^{10}$Stahl et al. 2008,  $^{11}$Wade et al. 2006,
$^{12}$Grunhut et al. 2009, $^{13}$Grunhut et al. 2012,
$^{14}$Wade et al. 2012
\end{table*}

\citet{leone04} argued that as per the standard radio emission gyrosynchrotron models, emission at a given frequency should be
emitted in a well-localised torus above the magnetic
poles. They examined the radio spectral energy distributions (SEDs) of five
 magnetic B-type stars in the range from 1 GHz to 87.7-GHz.
They concluded that the observed slopes of the radio spectra and
 the absence of millimetre emission are not generally compatible  
 with this model, and suggested a cut-off frequency of the radio emission. 
 
\citet{trigilio00} detected rapid,
intense, narrow-band and highly polarized radio bursts from the late B-type star CU Vir, which they
hypothesised to be Electron Cyclotron Maser Emission (ECME) at 1.4 GHz. Such emission is expected to occur principally at low frequencies 
\citep[$<1.5$ GHz, ][]{trigilio00,leto12}. However, to date, few such low frequency 
observations have been obtained, and generally they correspond to null results \citep[e.g.][]{george12}.

Clearly, a variety of diverse investigations of the radio emission of magnetic A- and B-type stars have been carried out, leading to an array of disconnected results. However, no similar studies have yet taken place for hotter magnetic O- and B-type stars. In an attempt to extend and homogenize the study of the physics of radio magnetospheres of hot stars, we have initiated a systematic survey of the radio emission properties of the known magnetic B- and O-type stars. Our survey includes observations at both high frequency (currently being obtained with the VLA) and at low frequency (currently being obtained with the GMRT).

In this first paper we describe a pilot study of 8 established magnetic
O- and B-type stars, observed with the GMRT in the 1390 MHz (20 cm) and
610 MHz (50 cm) bands,
 to characterize the radio emission properties of their magnetospheres at
very low frequencies, study their radio SEDs to understand magnetospheric properties
and probe mass loss from the star in the case of dominant free-free absorption.

In \S\ref{observations} we describe the  observations and data analysis. In \S\ref{results} we carry out
a detailed analysis and look for variability of the detected stars within the 
observing period, and in \S\ref{freefree} we discuss
the stars for which no radio emission was detected. In \S\ref{discussion} we discuss our results and the potential of low
frequency radio observations for providing new constraints on the physics of hot star magnetospheres.

\begin{table*}
\scriptsize
 \centering
  \caption{GMRT observations of magnetic B and O type stars\label{tab:gmrt}. }
  \begin{tabular}{@{}lllllllllll@{}}
  \hline
Star &  Obs Date &
Mean &
Mean & Flux density & rms &
 Obs Date  & Mean &
Mean & Flux density & rms \\
&   1.4 GHz&
HJD
 &
 Phase & F$_{20}$ (mJy) & $\mu$Jy & 0.6 GHz
 & HJD &  Phase
& $F_{50}$ (mJy)& $\mu$Jy \\
\hline
HD 215441 &   01-Nov-13 & 2456598.262$\pm$ 0.050
& 0.93&$1.49\pm0.10$  &46 &
 24-Oct-13 & 	2456590.168$\pm$	0.044 &
0.08 & $0.98\pm0.10$ & 56\\
HD 37479 &   31-Oct-13&	2456597.433$\pm$	0.170
& 0.09& $2.00\pm0.11$ & 57 &
 24-Oct-13 &	2456590.402$\pm$	0.044
& 0.18& $1.25\pm0.28$ & 51 \\
HD 37017 &   31-Oct-13&	2456597.422$\pm$	0.131
& 0.94 & $1.59\pm0.12$ & 53&
 24-Oct-13 &	2456590.430$\pm$	0.040
& 0.18& $0.59\pm0.32$ & 153 \\
HD 36485 &   27-Oct-13&	2456593.378$\pm$	0.050
& 0.94& $1.30\pm0.10$ & 54&
 26-Oct-13 &	2456591.494$\pm$	0.076
& 0.66& $0.59\pm0.17$ & 88 \\
HD 133880 &   24-Jan-14&	2456681.616$\pm$	0.055
& 0.84& $14.45\pm0.11$ & 49&
 14-Jan-14 &	2456671.590$\pm$	0.048
& 0.42& $2.04\pm0.15$ & 66 \\
HD 37022 &   31-Oct-13&	2456597.476	
& 0.4& $<4.95$ & 1650&
$\ldots$ & $\ldots$& $\ldots$ & $\ldots$ & $\ldots$ \\
HD 57682 &   01-Nov-13&	2456598.370	 & 0.13& $<0.16$ & 52&
$\ldots$ & $\ldots$& $\ldots$ & $\ldots$ & $\ldots$ \\
NGC 1624-2 &  02-Nov-13&	2456598.487	& 1.0& $<0.21$ & 71&
$\ldots$ & $\ldots$ &$\ldots$ & $\ldots$& $\ldots$ \\
\hline
\end{tabular}
\end{table*}

\section{Observations}
\label{observations}

The GMRT observations of 8 magnetic O- and B-type stars  were taken between 2013 October 24 to 2014
January 24 under cycle 25 in the 1390 (L-band) and 610 MHz bands.
Table \ref{tab:stars} gives details of the observed stars.
Out of these,
5 B-type stars were observed in both the 1390 and 610
MHz bands, while the 3 O-type stars were observed only in the 1390 MHz band. 
Each observation
was of approximately  3 hours in duration and a total of 26-29 good antennae
were used. For all observations, visibilities were recorded for two circular polarizations (RR and LL) with 
a bandwidth 32 MHz, divided into 256 frequency channels,  
and 16 s integration time. 
Calibrator sources were used to remove the effect of variation of the
instrumental and other non-astronomical factors in the measurements.
3C48, 3C286 and 3C147 were used as flux calibrators in various observations.
Flux calibrators were observed for 10--15 minutes, either in the beginning or towards the end of
each observation. Flux calibrators were also used
for bandpass calibration. 
Phase calibrators were chosen from the VLA calibrator
manual such that they are located within 15 degrees of the target star. They were observed more
frequently. This is 
important not only for the tracking of instrumental phase and
gain drifts and atmospheric and ionospheric gain and phase variations
but also for monitoring the quality and sensitivity of
the data and for spotting occasional gain and phase jumps.
 In the 1390 MHz band, the phase
calibrator observations were made approximately every 25--30 minutes for
5 minutes duration, whereas in the 610 MHz band, the 6-minute phase calibrator scans were obtained every 35--40 minutes. 

A FLAGging and CALibration (FLAGCAL) software pipeline developed for
automatic flagging and calibration of the GMRT data \citep{jayaram}
was used to flag and calibrate the GMRT data. In 70\% of the cases, the
results were satisfactory and only  minor additional flagging was needed after the
FLAGCAL. However, in some cases, the data needed more significant flagging
and recalibration. In such cases
 the flagging was done manually and a flag file was created. The
FLAGCAL was then run again using the manual flag file. The calibrated data were 
 inspected to determine  the presence of any spurious signature.
The above step was repeated until satisfactory calibration was achieved.
The flagged and calibrated visibility data were used to make continuum images using the standard 
tasks in the Astronomical Image Processing System (AIPS). 
To avoid bandwidth smearing in the continuum image, the total bandwidth was divided into 6 sub-bands in 
the 1390 MHz observations, and in 10 sub-bands in the 610 MHz observations.
Because of the large field of view (FoV) of the GMRT,
 the three-dimensional (3D) imaging feature in the AIPS task `IMAGR' was used
 in which the entire FoV is divided into 
multiple subfields (facets) and each of which was
 imaged separately. For the 1390 MHz image, the total FoV was divided in 19 subfields, and in the 610 MHz band the FoV
was divided into 37 subfields.
The presence of a large number of bright sources in the FoV of various stars allowed us to carry out 
self-calibration to improve the complex gains. This reduces the errors 
from temporal variations in the 
system gain, and spatial and temporal variations 
in the ionospheric properties. 
After 3 rounds of phase self-calibration, the clean component
model was subtracted from the UV data to identify the residual bad data.
Some more flagging was performed and 2 more rounds of phase self-calibration
were run. A final round of amplitude and phase self-calibration was
also performed in each dataset.
Since the phase variations occur 
on time-scales of a few minutes, the time interval for the 
phase self-calibration was chosen to be 1 minute, and 5 minutes
for the amplitude and phase self-calibration.
The map resolutions for the 1390 MHz images were around 2''--3'',
 and around 5''-6'' for the 610 MHz data.
In Table~\ref{tab:gmrt}, we report the values of the final flux density.
For non-detections, we quote $3\sigma$
upper limits.

\section{Analysis and results}
\label{results}

With our GMRT observations, we have detected 5 out of 8 stars.
The detected stars, HD 215441, HD 37479, HD 37017, HD 36485 and 
HD 133880, are B-type stars and have been detected in both the 1390 and 610 MHz bands.
 The O-type stars, HD 37022, HD 57682 and NGC 1624-2, were observed only in the 
1390 MHz band, and none of them are detected. 

Of the B stars, 4 stars (HD 37479, HD 37017, HD 36485 and HD 133880) are sufficiently rapid rotators that rotation contributes significantly to 
the support of their magnetospheres \citep[so-called `centrifugal magnetospheres'; ][]{petit13}. On the other hand, the O-type stars, as well as HD 215441, rotate slowly, and their magnetospheres receive no significant rotational support \citep[so-called `dynamical magnetospheres'; ][]{petit13}.

The observations are
summarized in Table \ref{tab:gmrt}. In this
section,  we give details of the detected B-type  stars, and explore
potential causes of non-detections in O-type stars.

\subsection{HD 215441}
HD 215441 (Babcock's star) is a cool magnetic B8-9p star. It exhibits 
no optical, UV or X-ray evidence of a magnetosphere, 
but was detected in the  5 GHz and 1.4 GHz bands  (flux density $\sim 1$~mJy) 
by \citet{linsky92}. Despite its low temperature, this star is noteworthy due to its extraordinarily strong magnetic field \citep{babcock1960}.

Magnetic O- and B-type stars show variability of many observational quantities. This variability is understood as rotational modulation in the context of the
oblique rotator model (ORM) \citep{stibbs50,wade11}, in which the dipolar
surface magnetic field is tilted with respect to the rotational axis.  Rotation of the star
results in strict periodic variability according to the stellar rotational period.

To phase our observations of HD 215441, we used the ephemeris of \citet{north1995}:

\begin{equation}
\label{north}
{\rm JD} = 2448733.714 + (9.487574\pm 0.000030)\cdot E.
\end{equation}

Radio emission is clearly detected in both the 1390 and 610 MHz bands
at phases 0.93 and 0.08, respectively, with a flux density 
somewhat larger than that reported at higher 
frequencies by \citet[][1.2--1.3~mJy]{linsky92}. In both bands, our observations cover less than 1\% of the rotational period; hence we are unable to evaluate any variability. The flux is somewhat ($\sim 15\%$) lower in the 610 MHz band. Whether this is due to the slope of the radio SED, or due to flux variability between phases 0.93 and 0.08, is not currently known.

\citet{linsky92} reported a flux of $1.07\pm0.09$~mJy at 1.4 GHz on July 2, 1987. According to Eq.~\ref{north}, this corresponds to phase 0.08, i.e. the same phase as our 610 MHz observation ($0.98\pm 0.1$~mJy). Comparing the two measurements, this implies an essentially flat spectral index of ($\alpha=\log(F_{1390}/F_{610})]/\log(1390/610)=+0.02\pm 0.03$) at this phase. This result appears inconsistent with the $\alpha=-0.26$ index derived by \citet{leone04} based on higher frequency observations. This suggests that the SED flattens at low frequency.

\subsection{HD\, 37479}

HD 37479 ($\sigma$~Ori E) is the prototypical B2 Vp star hosting a
 centrifugal magnetosphere. It exhibits strong and variable H$\alpha$
  emission that has been studied in some detail
   \citep[e.g.][]{2005ApJ...630L..81T,oksala12}. It was detected by
    \citet{linsky92} at 15,  5 and 1.4 GHz frequency 
bands using the VLA, and observed 
    at several phases. The flux density 
was observed to vary by a factor of about 2, from 2.4-3.1 mJy 
at 15 GHz, 2.8-3.9 mJy at 5 GHz, and 1.5-3.2 mJy at 1.4 GHz bands. 
They also detected variable circular polarization at 15 GHz. 

This star was also observed by \citet{lo93}, who detected
 5 GHz emission varying with the rotational period, with maxima coinciding with the magnetic extrema.

We obtained multiple GMRT observations in both bands. To phase the data, we use the ephemeris of \citet{townsend2010}, which takes into account the (slow) rotational braking of the star:

\begin{equation}
{\rm JD} = 2442778.8290+1.1908229\cdot E+1.44\times10^{-9}\cdot E^2.
\end{equation}

\begin{figure}
\centering
\includegraphics[width=0.45\textwidth]{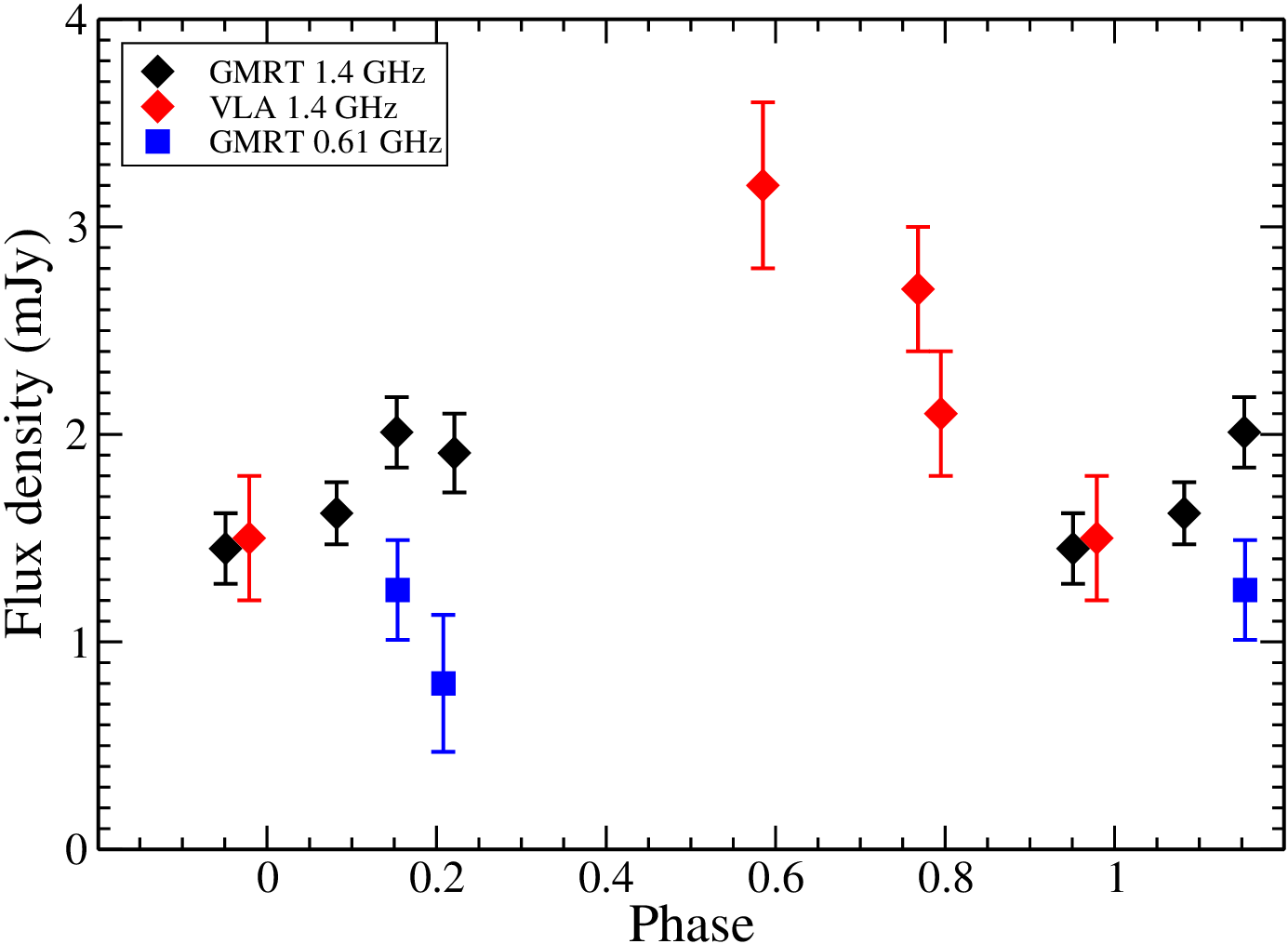}
\caption{GMRT and VLA 1.4 GHz measurements (black and red diamonds, 
respectively) and GMRT 610 MHz measurements (blue squares) of 
HD 37479 ($\sigma$~Ori E). The VLA measurements are taken from the archive.}
\label{sorie}
\end{figure}

Our observations of HD 37479 were obtained at phases 0.95, 0.08, 0.13 and 0.22 in the 1390 MHz band, and phases 0.15 and 0.21 in the 610 MHz band. We have supplemented our observations with the four 1.4 GHz measurements obtained by \citet{linsky92}.  The phased flux variation is shown in Fig.~\ref{sorie}. 

The combined 1390 MHz data suggest a roughly sinusoidal variation peaking at phase 0.5, with a peak flux of $\sim 3.2$~mJy. The 610 MHz measurements are both weaker than the minimum 1390 MHz measurement, but their phase coverage is too sparse to infer meaningful information about the phase variation in this band. 

\citet{lo93} proposed that the 5 GHz flux from this star exhibited a double-wave phase variation. This is not suggested to be the case at
 1.4 GHz band. Nevertheless, additional data will be required to draw more robust conclusions.

Two of the 1390/610 MHz GMRT measurements were obtained at 
similar phases. From these we infer an instantaneous ($\phi=0.15-0.2$) index of the SED at these frequencies of $\alpha= 0.11\pm 0.04$. 
This is consistent with $\alpha=0.12$ derived by  \citet{leone04}
 based on higher frequency observations.

\subsection{HD 37017}

HD 37017 is an early-type magnetic B2 Vp star that also exhibits H$\alpha$ emission due to the presence of a dense centrifugal magnetosphere \citep{leone2009,petit13}. This target is in fact a double-lined spectroscopic binary (SB2) system with an 18.65d orbital period containing the magnetic B star in addition to a late B dwarf \citep{leone2009}.

\citet{lo93} detected variable radio emission from HD 37017. Like $\sigma$ Ori E, the radio emission was found to vary with the rotational period, and magnetic extrema 
coincided with radio maxima. It was detected by \citet{linsky92} at 15, 5 and 1.4  GHz bands using the VLA, and observed at several phases. The flux 
density was observed to vary by a factor of about 2, from 0.9--2.1
 mJy at 15 GHz, 1.4--2.6 mJy at 5 GHz, and 1.5--2.4 mJy at 1.4 GHz bands.
\citet{leone04} studied HD 37017 in mm bands and examined the properties of the
radio spectrum in the 1.4--87.7 GHz range. 
They found the flux density to increase up to
a frequency of 22.5 GHz and then to decrease in the mm range. This indicates 
a possible cut-off frequency.

We obtained GMRT observations in the 1390 and 610 MHz  bands. We phased the data using the ephemeris of \citet{bolton1998}:

\begin{equation}
{\rm JD} = 2446010.3750 + (0.9011836\pm 0.000006)\cdot E
\end{equation}

\begin{figure}
\centering
\includegraphics[width=0.45\textwidth]{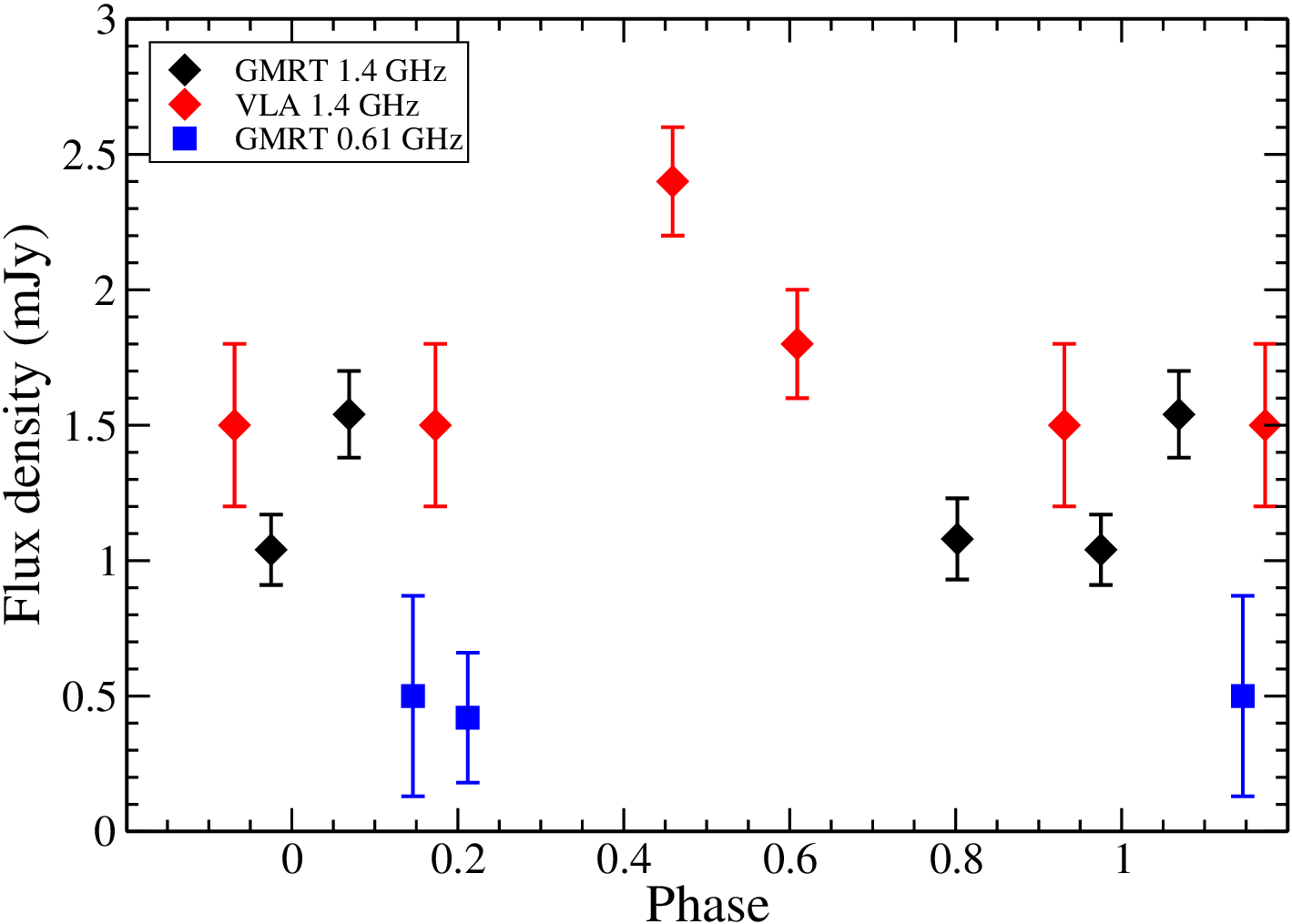}
\caption{GMRT and VLA 1.4 MHz measurements (black and red diamonds, respectively) and GMRT 610 MHz measurements (blue squares) of HD 37017.
The VLA measurements are taken from the archive.}
\label{hd37017}
\end{figure}

Our 1390 MHz observations were obtained at phases 0.07, 0.80 and 0.98, while our 610 MHz observations were obtained at phases 0.15 and 0.21. We supplemented these measurements with the 1420 MHz observations of \citet{linsky92}. The phased data are illustrated in Fig.~\ref{hd37017}.

\citet{lo93} described the 5 GHz flux variation as sinusoidal, with a maximum at phase 0.0 and a minimum at phase 0.5. The variation shown in Fig.~\ref{hd37017}, although weak and coarsely sampled, appears opposite to this description. This is entirely a consequence of the different ephemeris used by those authors and by us. The phasing relative to the longitudinal field variation is the same.

Again, our two 610 MHz observations exhibit lower flux than the lowest of the 7 measurements at 1390 MHz. One of the VLA 1.4 GHz observations was obtained at a phase similar to the GMRT 610 MHz data. From these we infer an instantaneous ($\phi=0.15-0.2$) index of the SED at these frequencies of $\alpha= 0.23\pm 0.1$. This is consistent with $\alpha=0.15$ derived by  \citet{leone04} based on higher frequency observations.

\subsection{HD 36485}

HD 36485 ($\delta$~Ori C)  is a another hot B3 Vp
 star hosting a centrifugal magnetosphere. 
The H$\alpha$  emission is variable on a timescale of a few hours. 
Notwithstanding its sharp spectral lines, the star is a rapid rotator 
with a period of about 1.5d, implying that it is viewed 
close to the rotational pole \citep{bohlender91,leone2009}.

$\delta$~Ori C was detected with the VLA at 5 GHz with a mean flux density of 0.95
mJy and showed a negative spectral index ($\alpha<-0.7$; $f_\nu \propto
\nu^{\alpha}$)  between 5 and 15 GHz bands according to \citet{drake87}, who concluded a non-thermal origin of the radio emission.
Their 3 measurements at this wavelength did not show any significant variability, with a measurement uncertainty of $\pm 0.1$ mJy.

We use the ephemeris of \citet{leone2009}:

\begin{equation}
{\rm JD} = 2448298.86 + (1.47775\pm 0.00004)\cdot E.
\end{equation}

In our GMRT observations, we detected this star in both the 1390 and 610 MHz bands. 
In our 3 hour observation at 1390 MHz covering phases 0.91-0.96, the flux density varied between $0.42\pm 0.24$ to $0.6\pm 0.14$ mJy. Although these measurements are formally consistent with a constant flux
density, the flux did monotonically increase during the observation, suggesting a real change.
In the 610 MHz band, the flux
density varied significantly, from $1.46\pm 0.16$ to $0.69\pm 0.13$ mJy between phases 0.62 to 0.71.

\subsection{HD 133880}

HD 133880 (HR Lup) is a cooler, rapidly rotating \citep[$P=0.88$~d; ][]{bailey12}
Bp star hosting 
one of the strongest known stellar magnetic fields. However, unlike most Ap/Bp stars, HD 133880 has a magnetic field topology that appears to be predominantly quadrupolar as opposed to dipolar \citep{landstreet90,bailey12}.
HD 133880 was previously observed with the Australia Telescope Compact Array (ATCA) at 5 and 8.5 GHz frequency bands simultaneously on 1995 February 12, 14 and 16, for around 10 hours each day.
\citet{lim96} had demonstrated that  radio flux and circular polarization of HD 133880 at both frequencies vary significantly and coherently according to the rotational period.
They reported that the emission shows broad peaks near the phases of the longitudinal field extrema.
The reanalysis of the data by \citet{bailey12} yielded the same result. 
They found flux variation to be complex, characterized by strong, broad maxima at phases 0.0 and 0.5 (i.e. the extrema of the longitudinal field), and sharper, somewhat weaker secondary extrema at quadrature phases (i.e. 0.25 and 0.75).

\citet{george12} observed this star with the GMRT in dual mode in the 610 and 240 MHz bands on December 5 and 7th, 
2009. They reported non-detections in all the observations. However, we extracted the data from the
GMRT archive and reanalysed it, and find detections on both days in the
610 MHz band. 
Furthermore, we also
find variability in each observation in this band. To re-confirm our detections of their data, we 
have analysed the data both in AIPS, as well as in Common Astronomy Software Applications
(CASA), separating out LL and RR Stokes, and we find consistent results.
However, in the 240 MHz band we obtained no detection in
either observation, consistent with \citet{george12}. The $3-\sigma$ upper limits in this band at the 
two epochs were
1.5 and 3.0 mJy, respectively. We also located three data sets at 1420 MHz in the 
VLA archive, obtained between 15--17 February 1995. We reanalyse these data and tabulate them in Table \ref{hd133880}, along with 
the flux density 
in various time intervals in both our new observations, as well as the reanalysed observations. 
 
To phase the data, we used the ephemeris of \citet{bailey12}:

\begin{equation}
{\rm JD} = 2445472.0 + (0.877476\pm 0.000009)\cdot E.
\end{equation}

\begin{figure}
\centering
\includegraphics[width=0.47\textwidth]{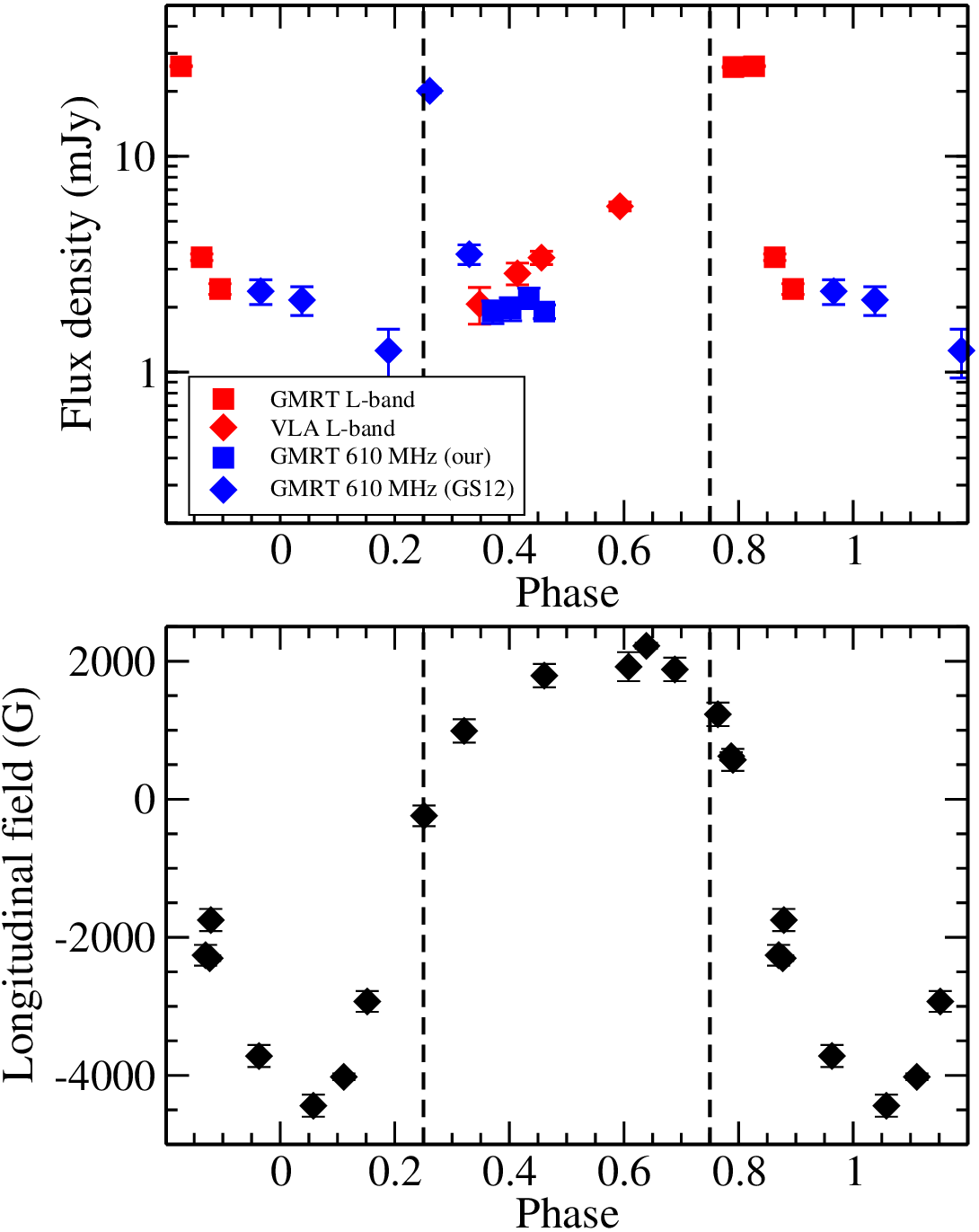}
\caption{{\em Upper frame -}\ Phase versus flux density plot of HD 133880 at 1.4 GHz (red color) and 610 MHz (blue color). Here red squares denote GMRT 1390 MHz measurements, red diamonds denote VLA 1.4 GHz measurements, blue squares denote GMRT 610 MHz observations made by us in 2014 January, whereas the blue diamonds are the 2009 December data of \citet{george12} reanalyzed by us. Note that unlike in Figs. 1 and 2, the vertical scale in this figure is logarithmic. {\em Lower frame -} Longitudinal magnetic field variation of HD 133880, from \citet{bailey12}. The vertical lines in both plots indicate phases 0.25 and 0.75.}
\label{fig:hd133880}
\end{figure}

The GMRT observations were made in the 1390 and 610 MHz bands. In Fig. \ref{fig:hd133880}, we plot the phased data in both bands. 
The flux variation of HD 133880 at both 1390 and 610 MHz is extraordinary. Unlike the other stars observed in this program (Table~ \ref{tab:variability}) that exhibit a flux variation of about a factor of 2, the 610 MHz flux of HD 133880 varies by a factor of 16 in our observations (and more than a factor of 12 at 1390 MHz). Even with the sparse phase coverage of our observations, it is clear that the maxima of both the 610 and 1390 MHz data occur at phases $\sim 0.25$ and 0.75, coincident with the phases of the minor maxima at 5 and 8 GHz described by \citet{bailey12}. 

Considering the important differences in the phase variations of the GMRT data and the ATCA data, we speculate that HD 133880 may exhibit maser emission in a manner similar to CU Vir \citep{trigilio00}. In CU Vir, while no coherent emission was found at 5 GHz, the 1.4 GHz emission was identified as ECME with a
basal flux of 2--3 mJy and then very large increments 
in the flux density at around phases 0.35-0.45 and 
0.75-0.85.
While, we do not yet have complete phase coverage for HD 133880, the observed features are highly reminiscent of those of CU Vir
(see Fig. \ref{fig:hd133880}). 

\begin{table*}
 \centering
 \begin{minipage}{100mm}
  \caption{Variability of HD 133880 \label{hd133880}.}
  \begin{tabular}{@{}lllllll@{}}
  \hline
 Freq & Telescope & Date & Mean & Phase
& Flux density & rms  \\
  GHz & &  & HJD
& & mJy & $\mu$Jy   \\
\hline
  1.388  & GMRT  & 24-Jan-14	& 2456681.572	$\pm$	0.012	&	0.791
& $25.90\pm0.14$ & 108 \\
           &     & 24-Jan-14 	&2456681.604	$\pm$	0.012	&	0.827
 & $26.14\pm0.12$ & 123 \\
             &   & 24-Jan-14 	& 2456681.636	$\pm$	0.012	&	0.863
& $3.41\pm0.12$ & 117 \\
              &  & 24-Jan-14 	&2456681.663	$\pm$	0.008	&	0.895
 & $2.43\pm0.14$ & 177 \\
  1.425 & VLA    & 15-Feb-95 	&2449764.040	$\pm$	0.007	&	0.348
 & $2.07\pm0.40$ & 396 \\
 & 		 & 15-Feb-95 	&2449764.098	$\pm$	0.006	&	0.414
 & $2.87\pm0.33$ & 331 \\
 &		 & 16-Feb-95 	& 2449765.013	$\pm$	0.006	&	0.456
& $3.39\pm0.24$ & 239 \\
 &		 & 17-Feb-95 	& 2449766.010	$\pm$	0.006	&	0.593
& $5.86\pm0.27$ & 267 \\     
  0.608 &GMRT  	 & 14-Jan-14 	& 2456671.552	$\pm$	0.010	&	0.371
& $1.91\pm0.23$ & 76 \\
            &    & 14-Jan-14 	&2456671.579	$\pm$	0.010	&	0.402
 & $1.97\pm0.24$ & 68 \\
     &           & 14-Jan-14 	&2456671.607	$\pm$	0.010	&	0.434
 & $2.21\pm0.24$ & 65 \\
           &     & 14-Jan-14  	&2456671.631	$\pm$	0.007	&	0.461
 & $1.91\pm0.14$ & 75 \\
    0.606 & GMRT & 05-Dec-09 	&2455170.712	$\pm$	0.026	&	0.966
 & $2.37\pm0.31$     & 311 \\
    &   	 & 05-Dec-09 	&2455170.776	$\pm$	0.027	&	0.038
 & $2.16\pm0.33$     & 328 \\
0.607&GMRT  	 & 07-Dec-09 	&2455172.663	$\pm$	0.027	&	0.189
 & $1.26\pm0.32$     & 218 \\
    &		 & 07-Dec-09 	&2455172.726	$\pm$	0.026	&	0.261
 & $20.08\pm0.40$     & 287 \\
     &		 & 07-Dec-09 	&2455172.787	$\pm$	0.024	&	0.330
 & $3.52\pm0.37$    & 347 \\               
\hline
\end{tabular}
\end{minipage}
\end{table*}

\begin{table*}
 \centering
  \caption{Variability of other stars in our sample. \label{tab:variability}}
  \begin{tabular}{@{}lclllrlll@{}}
  \hline
Star & Tel. & Freq & Date & Mean & Phase & Flux density & rms & Ref.\\
 & & GHz &  & HJD &
& mJy & $\mu$Jy &\\
\hline
HD215441 &  GMRT&  1.388   & 01 Nov 13	& 2456598.223	$\pm$	0.010	&	0.927	&  $1.43\pm0.18$& 76 & This paper\\
         &  &         & 01 Nov 13	& 2456598.255	$\pm$	0.010	&	0.931	&
 	$1.15\pm0.15$ & 70\\
         &       &    & 01 Nov 13	& 2456598.283	$\pm$	0.010	&	0.934	&
 $1.45\pm0.18$ & 74\\
         & &         & 01 Nov 13	& 2456598.306	$\pm$	0.005	&	0.936	&
 $1.20\pm0.19$ & 92\\
         & VLA & 1.4 &  02 Jul 87      & 2446979.250			&	0.078	& 
$1.07\pm0.09$ & $\ldots$ & \\         
HD215441 & GMRT&  0.608   & 24 Oct 13 & 2456590.136	$\pm$	0.012	&	0.075	&	
 $1.19\pm0.13$& 71 & This paper\\
         &      &     & 24 Oct 13&	2456590.168	$\pm$	0.012	&	0.078	&
  $1.18\pm0.13$ & 69\\
         &       &    & 24 Oct 13&	2456590.201	$\pm$	0.012	&	0.082	&
 $0.70\pm0.22$ & 108\\
HD37479  & GMRT&  1.388   & 31 Oct 13	&2456597.273	$\pm$	0.010	&	0.951	&
  $1.45\pm0.17$ & 87 & This paper\\
         &      &     & 31 Oct 13 	& 2456597.430	$\pm$	0.010	&	0.082	&
   $1.62\pm0.15$ & 84\\
         &       &    & 01 Nov 13 	& 2456597.514	$\pm$	0.010	&	0.153	&
  $2.01\pm0.17$ & 86\\
         &        &   & 01 Nov 13& 	2456597.595	$\pm$	0.008	&	0.221	&
  $1.91\pm0.19$ & 106\\
& VLA &        1.4     & 11 Mar 85 & 	2446136.467			&	0.585	&
  $3.2\pm0.4$ & $\ldots$ & \citet{linsky92}\\
& &             & 12 Mar 85 	& 2446136.685			&	0.768	&
  $2.7\pm0.3$ & $\ldots$ &\\
& &             & 16 Mar 85 		& 2446141.480			&	0.795	&
$2.1\pm0.3$ & $\ldots$ &\\
& &             & 17 Mar 85 		& 2446141.699			&	0.979	&
  $1.5\pm0.3$ & $\ldots$ &\\
HD37479  &  GMRT & 0.608   & 24 Oct 13 	&2456590.370	$\pm$	0.012	&	0.154	&
   $1.25\pm0.24$ & 134 & This paper\\
         &       &    & 24 Oct 13 	& 2456590.434	$\pm$	0.012	&	0.208	&
  $0.80\pm0.33$ & 158\\
HD37017  &  GMRT & 1.388    & 31 Oct 13 & 2456597.301	$\pm$	0.010	&	0.802	&
  $1.08\pm0.15$ & 74 & This paper\\
         &         &  & 31 Oct 13 	& 2456597.458	$\pm$	0.010	&	0.975	&
 $1.04\pm0.13$ & 66\\
         &          & & 01 Nov 13 	& 2456597.542	$\pm$	0.010	&	0.069	&
  $1.54\pm0.16$ & 78 \\
&VLA &        1.4     & 11 Mar 85 	& 2456136.488			&	0.459	&	
  $2.4\pm0.2$ &$\ldots$ 	& \citet{linsky92}\\
& &             & 12 Mar 85 		& 2456136.623			&	0.609	&
  $1.8\pm0.2$ &$\ldots$& \\
& &             & 16 Mar 85 		& 2456141.419			&	0.931	&
  $1.5\pm0.3$ &$\ldots$& \\
& &             & 17 Mar 85 		& 2456141.637			&	0.173	&
  $1.5\pm0.3$ &$\ldots$& \\
HD37017  &  GMRT & 0.608    & 24 Oct 13	& 2456590.402	$\pm$	0.012	&	0.146	&
   $0.50\pm0.37$ & 74& This paper\\
         &         &  & 24 Oct 13	& 2456590.462	$\pm$	0.008	&	0.212	&
  $0.42\pm0.24$ & 66\\
HD36485  &  GMRT&1.388    & 27 Oct 13 	& 2456593.338	$\pm$	0.010	&	0.910	&
 $0.43\pm0.17$ & 90\\
         &       &    & 27 Oct 13 	& 2456593.365	$\pm$	0.010	&	0.929	&
 $0.44\pm0.33$ & 85\\
         &        &   & 27 Oct 13 	& 2456593.393	$\pm$	0.010	&	0.947	&
 $0.60\pm0.14$ & 92\\
         &         &  & 27 Oct 13 	& 2456593.419	$\pm$	0.009	&	0.965	&
 $0.52\pm0.17$ & 78\\
  & VLA &        1.4     & 07 Jul 86 	&2446619.810			&	0.221	&
   $<1.2$ & 400& \citet{linsky92}\\      
HD36485  &  GMRT & 0.608    & 25 Oct 13 & 2456591.429	$\pm$	0.012	&	0.618	&
 $1.46\pm0.16$ & 108& This paper\\
         &    &       & 25 Oct 13 	& 2456591.461	$\pm$	0.012	&	0.640	&
 $1.26\pm0.18$ & 218\\
         &     &      & 26 Oct 13 	& 2456591.526	$\pm$	0.012	&	0.684	&
 $0.96\pm0.19$ & 87\\
         &      &     & 26 Oct 13 	& 2456591.558	$\pm$	0.012	&	0.706	&
 $0.69\pm0.13$ & 97\\
\hline
\end{tabular}
\end{table*}

\subsection{Free-free absorption in O star winds}
\label{freefree}

\citet{schnerr2007} carried out 5 and 1.4 GHz observations of 5 (non-magnetic)
O stars
with the Westerbork Synthesis Radio Telescope (WSRT). They detected 3 stars, for the
first time at 1.4 GHz band,  with one star
($\zeta $ Per) inferred to show non-thermal radio emission while two other ($\alpha$~Cam and $\lambda$~Cep) were inferred
to show thermal radio emission. In all cases, the observed flux was lower than their predicted
thermal flux, based on the mass loss rates inferred from the H$\alpha$ line profiles.
They explained this discrepancy as due to a higher influence of wind clumping on the formation of the H$\alpha$ line.

In our program we also observed 3 magnetic O-type stars at 1390 MHz. None of these targets were detected (see Table~\ref{tab:gmrt}). For one of these stars (HD 37022) our upper limit is quite poor because it resides in a region with extended
radio emission in Orion.

The high mass loss rates of O stars can make the wind optically thick due to free-free
absorption out to a very
large ``radio photosphere".
In such a case radio observations are unlikely to directly see the magnetospheric emission.
On the other hand, they may probe the global mass loss,
which is predicted to be quite heavily reduced for O stars with dynamical magnetospheres,
due to quenching of the mass
loss by the field and the resultant plasma in-fall \citep{uddoula02,uot08}.

We can estimate the radius of the free-free radio photosphere, under the assumption of a spherically-symmetric, non-magnetic wind,
 using Eq. (4) of \citet{torres11}:

\begin{equation}
\tau_{\rm ff}=5\times 10^3 \dot M^2 v_{\rm \infty}^{-2} f^{-1} \nu^{-2} T^{-3/2} R_{\rm ff}^{-3}
\end{equation}

\noindent where $\tau_{\rm ff}$ is the free-free optical depth, $\dot M$ is the mass-loss 
rate in units of $10^{-8}~M_\odot$/yr, $v_\infty$ is the wind terminal velocity in 
units of $10^8$ cm/s, $f$ is the clump volume filling
 factor, $\nu$ is the frequency of observation in GHz, $T$ is  the wind temperature in units of $10^5$~K, which we here for simplicity approximate with the effective temperature of the star, and $R_{\rm ff}$ is the distance from the star, in units of $3\times 10^{12}$ cm.

\begin{figure}
\centering
\includegraphics[width=0.45\textwidth]{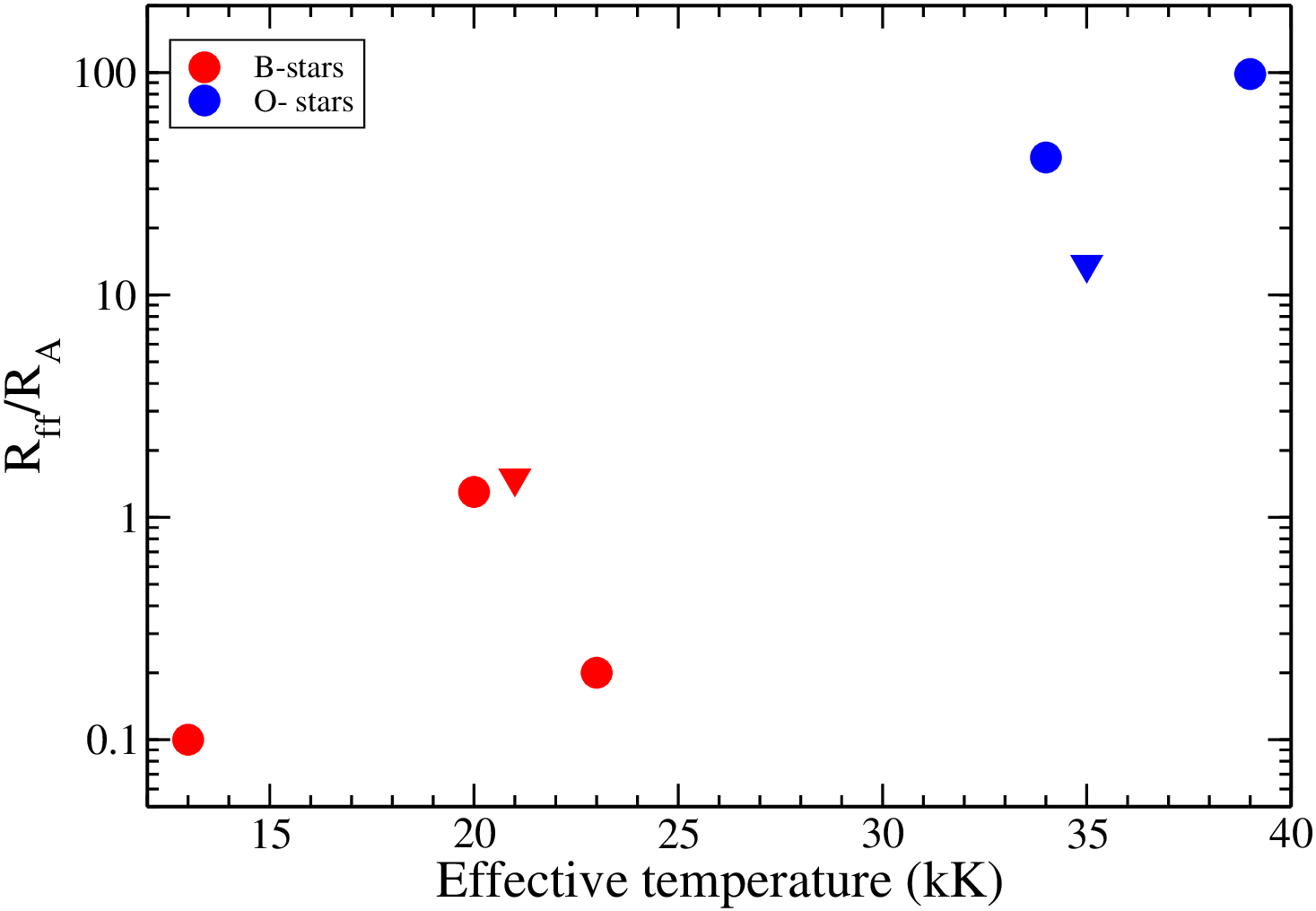}
\caption{Ratio of $R_{\rm ff}$ to $R_{\rm A}$ versus effective temperature for the observed sample (excluding HD 215441, for which $R_{\rm A}$ is unavailable). Here red circles and blue circles are for B- and O- type
stars. The inverted triangle indicate upper limits as listed in Table~\ref{tab:stars}. 
}
\label{ratio}
\end{figure}

We adopt a volume filling factor $f=1$, i.e. a smooth, unclumped radio emitting region, 
since it has been demonstrated \citep{puls06}
 that $f=1$ yields the best agreement with observations of thermal 
radio emission from hot stars when the \citet{vink00}
scaling law for mass-loss rates is assumed.
Assuming $\tau_{\rm ff}=1$, we have computed the position of the 
radio photosphere ($D$, in units of $R_*$) at 1390 MHz, using 
the wind parameters (computed using the Vink scaling) and 
physical parameters reported by \citet{petit13}, except for HD 133880, 
for which we use values reported by \citet{bailey12}. 
The results are summarized in Table~\ref{tab:stars} and illustrated in Fig.~\ref{ratio}.

If $R_{\rm ff}<R_{\rm A}$ we expect the magnetospheric emission to 
escape the wind and to be detectable (in principle). However, 
if $R_{\rm ff}>R_{\rm A}$, the magnetospheric emission will be hidden within 
the (thermal) radio photosphere of the wind.

In our sample, B stars
show ratios of $R_{\rm ff}/R_{\rm A}$ which are clustered around values 0.1--1.5, whereas 
all the O-type stars have this ratio clustered between 10--100. 
This is illustrated in Fig.~\ref{ratio}.
This convincingly demonstrates that the free-free absorption  of the magnetospheric 
radio emission of the B stars likely 
is negligible, and that the
principal cause of the null results for O stars is free-free 
scattering in the free streaming wind.

Current models of the magnetically confined winds of slowly rotating O stars imply that they
host ``dynamical magnetospheres". In this scenario, wind streamlines 
inside $R_{\rm A}$ in the magnetic equatorial region are constrained to 
follow closed field loops. After the wind stream reach the respective loop 
top and shock, the cooling plasma returns to the stellar surface on a relatively short (dynamical) timescale. Simulations \citep{uddoula02} predict that as a consequence of this confinement and fall-back, the global mass loss of O stars with dynamical magnetospheres should be significantly reduced, by up to a factor of 10--20 compared to similar non-magnetic stars.

In the context of free-free absorption, this reduction in mass loss has important implications for
 the location of the radio photosphere. The expected mass-loss rate can be estimated using equation 23 of 
 \citet{uot08}, leading to a reduction of $\dot M$ by a factor of more than 3 for HD 37022, 5 for HD 57682, and 15 for NGC 1624-2. If we reduce the mass loss rate in Eq. (6) by these factors, for HD 37022 and HD 57682 the ratio $R_{\rm ff}/R_{\rm A}$ is reduced to roughly 40, i.e. still well outside the Alfv\'en radius. However, the ratio for NGC 1624-2 is reduced to $\sim 2$. At shorter wavelengths (3-6 cm), this value becomes $<1$. Hence multiwavelength radio observations spanning the radio spectrum have the potential to provide novel constraints on predictions of mass-loss quenching of strongly magnetic O stars like NGC 1624-2.

\section{Summary and discussion}
\label{discussion}

We have studied 8 O- and B-type stars with the GMRT in the 1390 and 610 MHz bands.
Our 610 MHz detections of the 5 B type stars are the first 
detections of this class of objects at such low frequencies. This indicates that  
free-free absorption of their winds is not able to shadow their low frequency emission.
We also reanalysed archival observations of HD 133880 obtained in December 2009, for which \citet{george12} claimed non-detections. Surprisingly, we do detect the radio emission in both their observations allowing us to diagnose variability.

For  HD 37017 and HD 37479, we confirm weak variability of the 1390 MHz flux density, as reported by \citet{lo93} in 5 GHz band. The variability appears to be in phase with the variation of the longitudinal magnetic field. We do not find evidence of a double-wave 1390 MHz variation of HD 37479 \citep[as reported 
by][at 5 GHz]{lo93},
 although better phase coverage is necessary to robustly confirm this conclusion.

For the B star HD 36485, our observations suggest a relatively rapid change in the flux density in both bands in narrow phase ranges. This result needs to be confirmed by further observations. 

The cooler B star HD 133880 shows remarkable variability in both the 1390 and 610 MHz bands. Even with our limited phase coverage, the initial results suggest that the fluxes in both bands peak at phases coinciding with the minor peaks obtained from ATCA data \citep{bailey12}. We propose that this behaviour may be a consequence of ECME, first observed from the B8p star CU Vir by \citet{trigilio00}. The phenomenon, only observable at low radio frequencies, strongly 
motivates immediate monitoring of this target.

There were no detections obtained for any of the  three magnetic O stars in the 1390 MHz 
band. Using the scaling law of  \citet{torres11}, 
we are able to conclude that this null result is in general agreement with the expected free-free optical
depth of dense winds of O-type stars. We also conclude that radio observations at higher frequencies will be capable of testing predictions of mass loss quenching of strongly magnetic O stars like NGC 1624-2.

\citet{leone04} commented that the radio spectra of HD 142301 and HD 215441 (the two stars in their sample with the strongest magnetic
 fields) exhibited a sharp drop in flux density between 15 and 22.5 
 GHz. They concluded that this suggests there is a cut-off frequency 
 in this range. On the other hand, the magnetic chemically peculiar 
 stars with relatively weak magnetic fields showed flux densities 
 that certainly decrease from the cm to the mm range. 
In our sample,  we were able to measure spectral indices between 1390
 and 610 MHz for 3 stars in common with the sample of \citet{leone04}.
 For two stars (HD 37017 and HD 37479), the indices were similar to those obtained at higher frequencies, suggesting a smooth decrement 
 of the flux density. On the other hand, for HD 215441 (the nondegenerate star with the strongest known magnetic field), we obtained a
  spectral index that differed strongly from that derived by
   \citet{leone04}. This implies an equally significant change in the spectral index at low frequencies, and deserves further investigation.

The results presented in this paper represent a first feasibility study
 of the detection of low frequency emission from magnetic O- and B-type
  stars. Having demonstrated the general properties of a range of
   objects, we are now carrying out a larger survey to observe all known
    magnetic OB stars in radio bands from low to high frequencies. We
     are also performing monitoring of individual magnetic stars, in
      order to better understand the rotational variation of their
       fluxes and spectral energy distributions.

\section*{Acknowledgments}

We thank the staff of the GMRT that made these
observations possible. GMRT is run by the National
Centre for Radio Astrophysics of the Tata Institute of Fundamental Research.
AIPS is produced and maintained by the National Radio Astronomy
Observatory, a facility of the National Science Foundation
operated under cooperative agreement by Associated Universities, Inc.
GAW acknowledges Discovery Grant support from the Natural Science
and Engineering Research Council (NSERC) of Canada.
AuD acknowledges support by NASA through Chandra Award number TM4-15001A and 16200111 and DHC for TM4-15001B issued by the Chandra X-ray Observatory Center which is operated by the Smithsonian Astrophysical Observatory for and behalf of NASA under contract NAS8- 03060.
AuD also acknowledges support for Program number HST-GO-13629.008-A provided by NASA through a grant from the Space Telescope Science Institute, which is operated by the Association of Universities for Research in Astronomy, Incorporated, under NASA contract NAS5-26555.

\setlength{\labelwidth}{0pt}

\end{document}